\documentclass[
    ,final            
  ]
  {aipproc}

\layoutstyle{6x9}

\begin{document}
\newcommand{\ds}{\displaystyle}
\newcommand{\sss}{\scriptscriptstyle}
\def \eps{\varepsilon}
\def \deg{^{\circ}}
\newcommand{\shat}{\ensuremath{{\hat s}}}
\newcommand{\limit}{\ensuremath{_{\scriptscriptstyle \mathrm{lim}}}}
\newcommand{\lum}{\ensuremath{{\mathcal L}}}
\newcommand{\ubar}{\ensuremath{\bar{u}}}
\newcommand{\dbar}{\ensuremath{\bar{d}}}
\newcommand{\ebar}{\ensuremath{\bar{e}}}
\newcommand{\mubar}{\ensuremath{\bar\mu}}
\newcommand{\nubar}{\ensuremath{\bar\nu}}
\newcommand{\ra}{\rightarrow}
\def\prop{\sim}
\def\gapf{\ensuremath{f(\Delta\eta)}}
\def\lsim{\,\lower.25ex\hbox{$\scriptstyle\sim$}\kern-1.30ex%
\raise 0.55ex\hbox{$\scriptstyle <$}\,}
\newcommand {\pom} {I\!\!P}
\newcommand {\spom} {\mbox{\footnotesize{$\pom$}}}
\newcommand {\pomsub} {{\scriptscriptstyle \pom}}
\newcommand {\reg} {I\!\!R}
\newcommand {\regsub} {{\scriptscriptstyle \reg}}
\newcommand {\ppp} {\pom\pom\pom}
\newcommand {\ppr} {\pom\pom\reg}
\newcommand {\xpom} {x_{\pomsub}}
\newcommand {\apom} {\alpha_{\pomsub}}
\newcommand {\areg} {\alpha_{\regsub}}
\newcommand {\aprime} {\alpha^\prime_\pomsub}
\newcommand {\deta} {\Delta\eta}
\newcommand {\aveapom} {\bar{\alpha}_\pomsub}
\newcommand{\rfour}{\mbox{$r^{04}_{00}$}}
\newcommand{\rfive}{\mbox{$r5_{00}$}}
\newcommand{\rfivecomb}{\mbox{$r5_{00} + 2 r5_{11}$}}
\newcommand{\ronecomb}{\mbox{$r1_{00} + 2 r1_{11}$}}
\newcommand{\tprim}{\mbox{$t^\prime$}}
\newcommand{\tlc}{\mbox{$t$}}
\newcommand{\mxtwo}{M2_X}
\newcommand{\xbj}{x_{\mathrm{Bj}}}
\newcommand {\xl} {x_L}
\newcommand{\yjb}{y_{\scriptscriptstyle JB}}
\newcommand {\mxlps} {M_{X}^{\mathrm{LPS}}}
\newcommand{\gp}{\gamma p}
\newcommand{\gvp}{\gamma^* p}
\newcommand{\etamax}{\eta_{\mathrm{max}}}
\newcommand{\dargthree}{\xpom\,,\beta,\,Q^2}
\newcommand{\dargfour}{\xpom,\,t\,,\beta,\,Q^2}
\newcommand{\ftwod}{F_2^D}
\newcommand{\ftwodthree}{F_2^{D(3)}}
\newcommand{\ftwodfour}{F_2^{D(4)}}
\newcommand{\fldthree}{F_L^{D(3)}}
\newcommand{\fldfour}{F_L^{D(4)}}
\newcommand{\srdthree}{\sigma_r^{D(3)}}
\newcommand{\srdfour}{\sigma_r^{D(4)}}
\newcommand{\fonedthree}{F_1^{D(3)}}
\newcommand{\fonedfour}{F_1^{D(4)}}
\newcommand{\ftwodthreearg}{F_2^{D(3)}\,(\dargthree)}
\newcommand{\ftwodfourarg}{F_2^{D(4)}\,(\dargfour)}
\newcommand{\fldthreearg}{F_L^{D(3)}\,(\dargthree)}
\newcommand{\fldfourarg}{F_L^{D(4)}\,(\dargfour)}
\newcommand{\fonedthreearg}{F_1^{D(3)}\,(\dargthree)}
\newcommand{\fonedfourarg}{F_1^{D(4)}\,(\dargfour)}
\newcommand{\srdthreearg}{\sigma_r^{D(3)}\,(\dargthree)}
\newcommand{\srdfourarg}{\sigma_r^{D(4)}\,(\dargfour)}
\newcommand{\ftwopom}{F_2^{\pomsub}}
\newcommand{\flpom}{F_L^{\pomsub}}
\newcommand{\fonepom}{F_1^{\pomsub}}
\newcommand{\ftwopomarg}{\ftwopom\,(\beta,\,Q^2)}
\newcommand{\flpomarg}{\flpom\,(\beta,\,Q^2)}
\newcommand{\fonepomarg}{\fonepom\,(\beta,\,Q^2)}
\newcommand{\pomflux}{f_{\pomsub/p}}
\newcommand{\pomfluxargs}{\pomflux(\xpom,\,t)}
\newcommand{\pomfluxarg}{\pomflux(\xpom)}
\def\cts{\cos\theta^{\ast}}
\def\xgobs{x_\gamma^{\scriptscriptstyle \mathrm{OBS}}}
\def\xpobs{x_p^{\scriptscriptstyle \mathrm{OBS}}}
\def\xg{x_\gamma}
\def\ETJ{E_T^{\mathrm{jet}}}
\def\ETAJ{\eta^{\mathrm{jet}}}

\newcommand{\stat}{\mathrm{stat}}
\newcommand{\syst}{\mathrm{syst}}
\newcommand{\model}{\mathrm{model}}

\newcommand\units[1]{\,\mathrm{#1}} 
\newcommand\fig[1]{fig.\,\ref{fig:#1}} 
\newcommand\Fig[1]{Fig.\,\ref{fig:#1}}
\newcommand\tab[1]{table \ref{table:#1}}
\newcommand\Tab[1]{Table \ref{table:#1}}
\newcommand\qq[1]{Eq.\,(\ref{qq:#1})}
\newcommand\sect[1]{\S\ref{sect:#1}}

\def\ap#1#2#3   {{\em Ann. Phys. (NY)} {\bf#1} (#2) #3}   
\def\apj#1#2#3  {{\em Astrophys. J.} {\bf#1} (#2) #3} 
\def\apjl#1#2#3 {{\em Astrophys. J. Lett.} {\bf#1} (#2) #3}
\def\app#1#2#3  {{\em Acta. Phys. Pol.} {\bf#1} (#2) #3}
\def\ar#1#2#3   {{\em Ann. Rev. Nucl. Part. Sci.} {\bf#1} (#2) #3}
\def\cpc#1#2#3  {{\em Computer Phys. Comm.} {\bf#1} (#2) #3}
\def\epj#1#2#3  {{\em Eur. Phys. J.} {\bf#1} (#2) #3}
\def\err#1#2#3  {{\it Erratum} {\bf#1} (#2) #3}
\def\ib#1#2#3   {{\it ibid.} {\bf#1} (#2) #3}
\def\jmp#1#2#3  {{\em J. Math. Phys.} {\bf#1} (#2) #3}
\def\ijmp#1#2#3 {{\em Int. J. Mod. Phys.} {\bf#1} (#2) #3}
\def\jetp#1#2#3 {{\em JETP Lett.} {\bf#1} (#2) #3}
\def\jpg#1#2#3  {{\em J. Phys. G.} {\bf#1} (#2) #3}
\def\mpl#1#2#3  {{\em Mod. Phys. Lett.} {\bf#1} (#2) #3}
\def\nat#1#2#3  {{\em Nature (London)} {\bf#1} (#2) #3}
\def\nc#1#2#3   {{\em Nuovo Cim.} {\bf#1} (#2) #3}
\def\nim#1#2#3  {{\em Nucl. Instr. Meth.} {\bf#1} (#2) #3}
\def\np#1#2#3   {{\em Nucl. Phys.} {\bf#1} (#2) #3}
\def\pcps#1#2#3 {{\em Proc. Cam. Phil. Soc.} {\bf#1} (#2) #3}
\def\pl#1#2#3   {{\em Phys. Lett.} {\bf#1} (#2) #3}
\def\prep#1#2#3 {{\em Phys. Rep.} {\bf#1} (#2) #3}
\def\prev#1#2#3 {{\em Phys. Rev.} {\bf#1} (#2) #3}
\def\prl#1#2#3  {{\em Phys. Rev. Lett.} {\bf#1} (#2) #3}
\def\prs#1#2#3  {{\em Proc. Roy. Soc.} {\bf#1} (#2) #3}
\def\ptp#1#2#3  {{\em Prog. Th. Phys.} {\bf#1} (#2) #3}
\def\ps#1#2#3   {{\em Physica Scripta} {\bf#1} (#2) #3}
\def\rmp#1#2#3  {{\em Rev. Mod. Phys.} {\bf#1} (#2) #3}
\def\rpp#1#2#3  {{\em Rep. Prog. Phys.} {\bf#1} (#2) #3}
\def\sjnp#1#2#3 {{\em Sov. J. Nucl. Phys.} {\bf#1} (#2) #3}
\def\spj#1#2#3  {{\em Sov. Phys. JEPT} {\bf#1} (#2) #3}
\def\spu#1#2#3  {{\em Sov. Phys.-Usp.} {\bf#1} (#2) #3}
\def\zp#1#2#3   {{\em Zeit. Phys.} {\bf#1} (#2) #3}

\def\zn#1       {{\em Zeus Note} {\bf#1}}
\def\dn#1       {{\em Desy Note} {\bf#1}}
\def\hepph#1    {\mbox{arXiv:hep-ph/#1\/} }
\newcommand{\mx}{\ensuremath{M_{X}}}
\newcommand{\half}{\ensuremath{\frac{1}{2}}}
\newcommand{\gs}{\gamma^{\ast}}
\newcommand{\as}{\alpha_{S}}
\newcommand{\chipdf}{\frac{\chi^2}{\mbox{{\tiny d.o.f.}}}}
\newcommand{\der}[2]{\frac{d#1}{d#2}}
\newcommand{\pder}[2]{\frac{\partial#1}{\partial#2}}

\newcommand{\EPS}[3]{
  \noindent
  \begin{figure*}[htb]
  \begin{center}
  \epsfig{figure=#1.eps,#3}
  \caption{{\em #2}}
  \label{fig:#1}
  \end{center}
  \end{figure*}
}

\newcommand{\EPScp}[3]{
  \EPS{#1}{#2}{#3}
  \clearpage
}

\newcommand{\lpage}{\enlargethispage{\baselineskip}}
\newcommand{\spage}{\enlargethispage{-\baselineskip}}
\newcommand{\error}[1] {{\ensuremath{\pm #1}}}

\title{The Pomeron Structure and Diffractive Parton Distributions}


\author{Halina Abramowicz\footnote{also at Max Planck Institute, Munich, Germany, Alexander von Humboldt Research Award.}, Michael Groys and Aharon Levy}{
  address={School of Physics and Astronomy, Raymond and Beverly Sackler Faculty of Exact Sciences, Tel Aviv University, Tel Aviv, Israel}
}



\begin{abstract}
Measurements of the diffractive structure
  function, $\ftwod$, of the proton at HERA are used    to extract the
  partonic structure of the Pomeron.  Regge Factorization is tested
  and is found to describe well the existing data within the selected
  kinematic range.  The analysis is based on the next to leading order
  QCD evolution equations.  The results obtained from various data
  sets are compared. An analysis of the uncertainties in determining
  the parton distributions is provided. The probability of diffraction
  is calculated using the obtained results.
\end{abstract}

\maketitle


\section{INTRODUCTION}

In the last 10 years a large amount of diffractive data was
accumulated at the HERA collider~\cite{zeus-diff,h1-diff,lps-diff}.
There are three methods used at HERA to select diffractive events.
One uses the Leading Proton Spectrometer (LPS)~\cite{lps-diff} to
detect the scattered proton and by choosing the kinematic region where
the scattered proton looses very little of its initial longitudinal
energy, it ensures that the event was diffractive. A second
method~\cite{h1-diff} simply requests a large rapidity gap (LRG) in
the event and fits the data to contributions coming from Pomeron and
Reggeon exchange. The third method~\cite{zeus-diff} relies on the
distribution of the mass of the hadronic system seen in the detector,
$M_X$, to isolate diffractive events and makes use of the Forward Plug
Calorimeter (FPC) to maximize the phase space coverage. We will refer
to these three as ZEUS LPS, H1 and ZUES FPC methods.

The experiments~\cite{ZEUS,H1,LPS} provide sets of results for
inclusive diffractive structure function, $\xpom \ftwodthree$, in
different regions of phase space. In extracting the initial Pomeron
parton distribution functions (pdfs), the data are fitted assuming the
validity of Regge factorization.

In the present study, Regge factorization is tested.  New fits, based
on a NLO QCD analysis, are provided and include the contribution of
the longitudinal structure function.  The obtained PDFs are
systematically analyzed.  A comparison of the different experimental
data sets is provided.  Additional quantities derived from the fit
results are also presented.

In order to make sure that diffractive processes are selected, a cut
of $\xpom <$ 0.01 was performed, where $\xpom$ is the fraction of the
proton momentum carried by the Pomeron. It was shown~\cite{GKS} that
this cuts ensures the dominance of Pomeron exchange. In addition, a
cut of $Q^2 >$ 3 GeV$^2$ was performed on the exchanged photon
virtuality for applying the NLO analysis.  Finally, a cut on $M_X >$ 2
GeV was used so as to exclude the light vector meson production.

\section{Regge factorization}

The Regge Factorization assumption  can be reduced to the
following,
\begin{eqnarray}
\ftwodfourarg &=& f(\xpom,t) \cdot F(\beta,Q^2), \label{qq:rf_assum}
\end{eqnarray}
where $f(\xpom,t)$ represents the Pomeron flux which is assumed to be
independent of $\beta$ and $Q^2$ and $F(\beta,Q^2)$ represents the
Pomeron structure and is $\beta$ and $Q^2$ dependent.  In order to
test this assumption, we check whether the flux $f(\xpom,t)$ is indeed
independent of $\beta$ and $Q^2$ on the basis of the available
experimental data.

The flux is assumed to have a form $\prop \xpom^{-A}$ (after
integrating over $t$ which is not measured in the data) . A fit of this
form to the data was performed in different $Q^2$ intervals, for the
whole $\beta$ range, and for different $\beta$ intervals for the whole
$Q^2$ range.

\begin{figure}[h]
  \includegraphics*[height=.15\textheight]{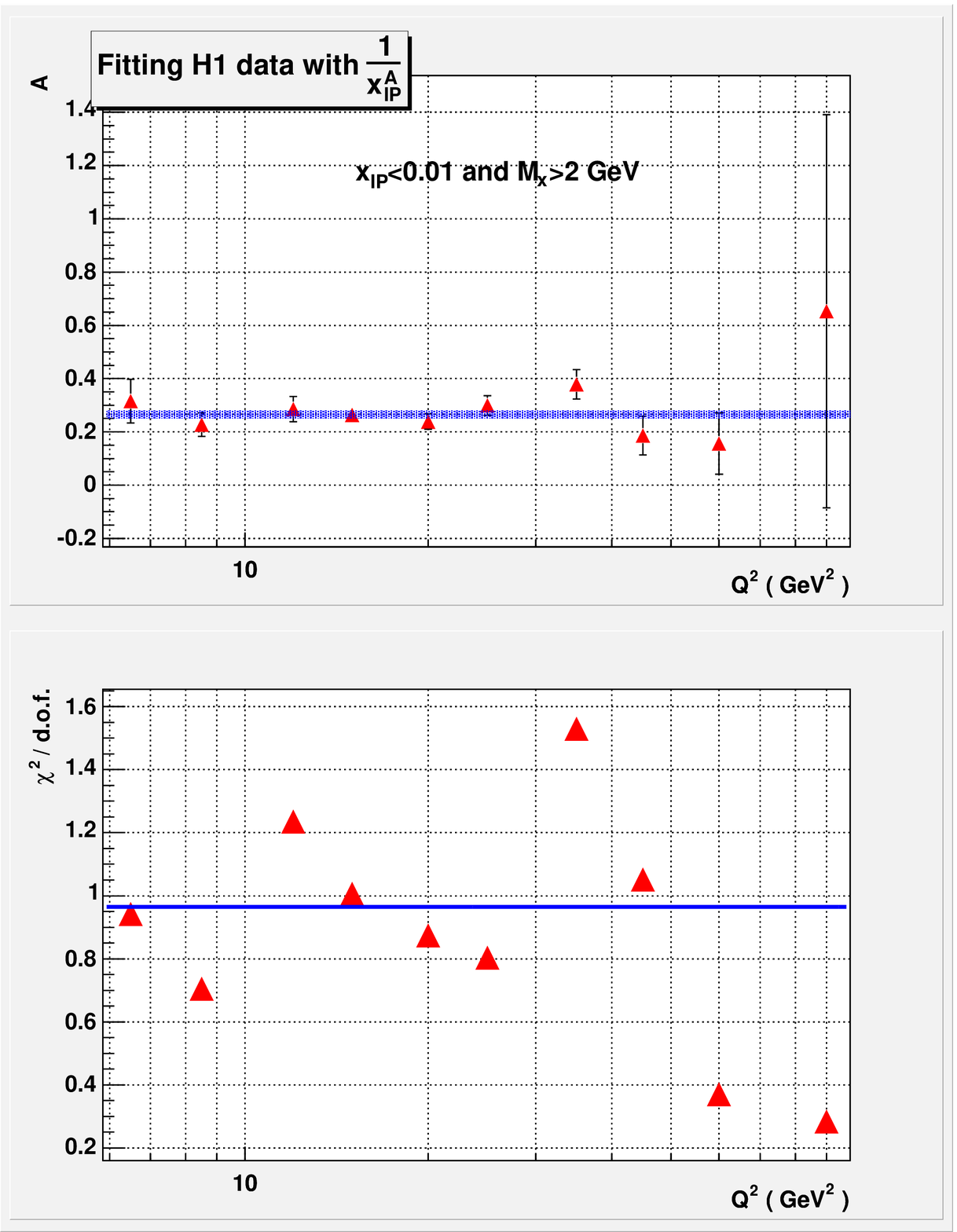}
  \includegraphics*[height=.15\textheight]{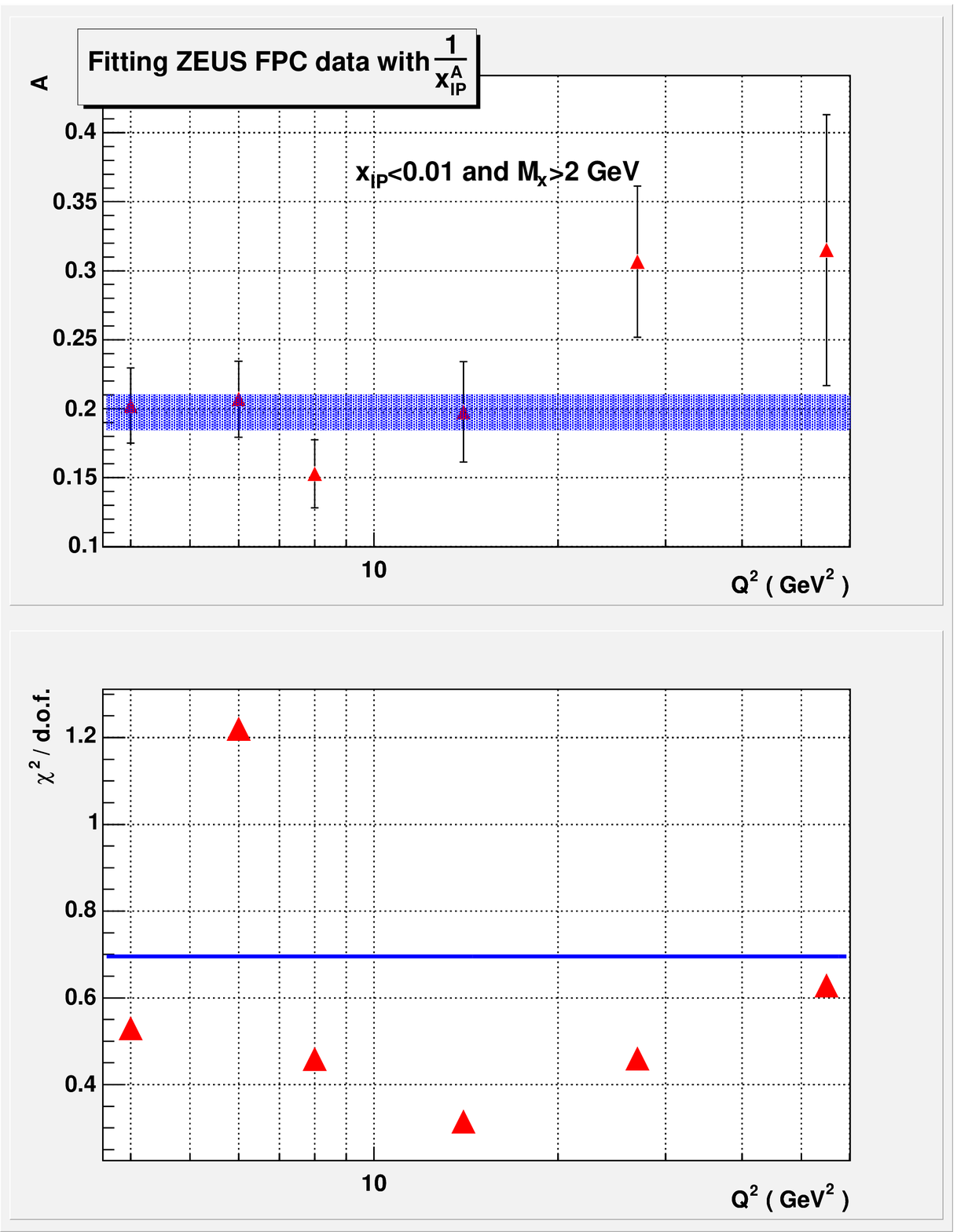}
  \includegraphics*[height=.15\textheight]{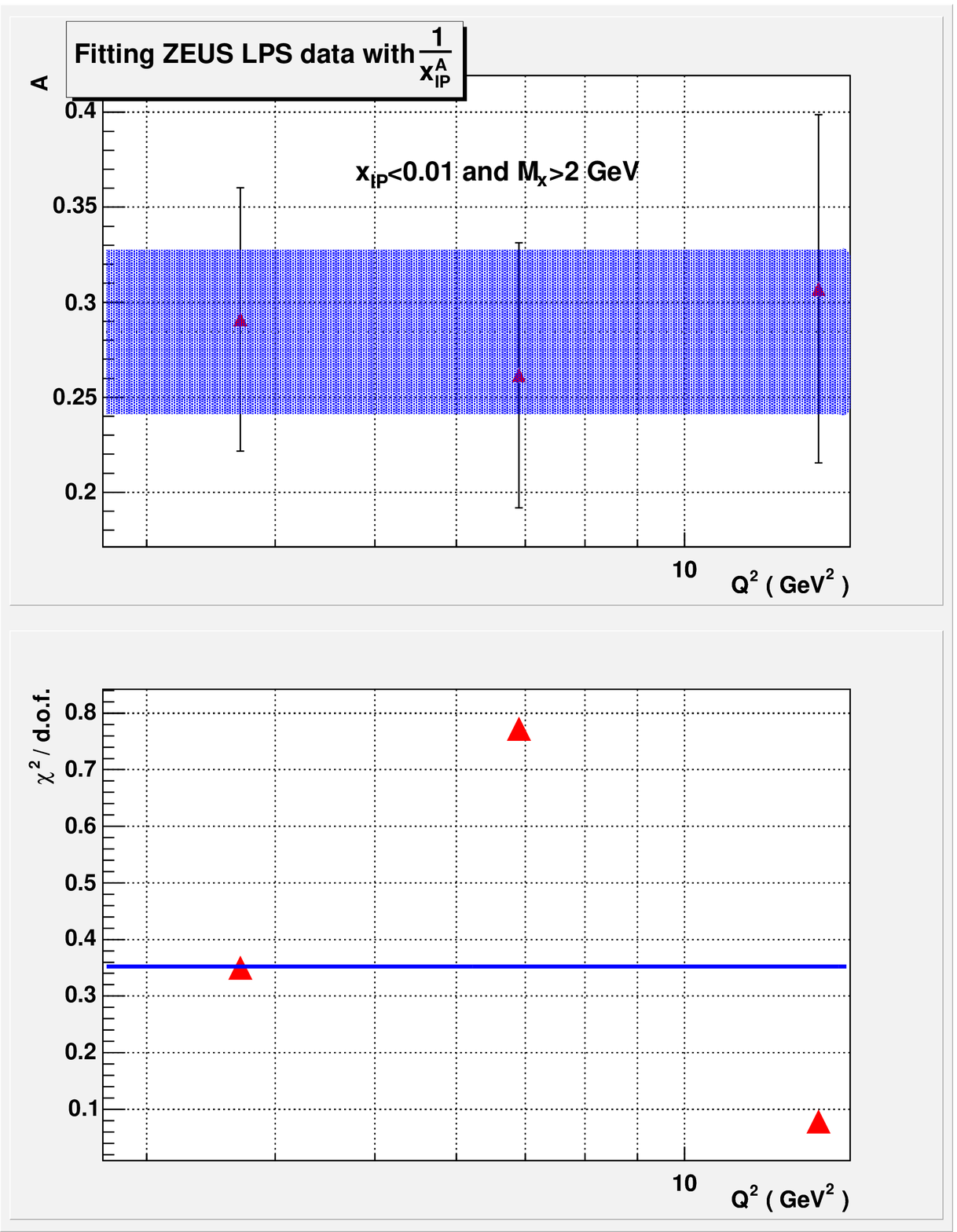}
  \caption{$A$ as a function of $Q^2$ for $\xpom <$ 0.001 and $M_X >$ 2 GeV, 
for the three data sets, as indicated in the figure. The line corresponds to 
a fit over the whole $Q^2$ region}
\label{fig:rf-q2}
\end{figure}

Figure~\ref{fig:rf-q2} shows the $Q^2$ dependence of the exponent $A$
for all three data sets, with the $\xpom$ and $M_X$ cuts as described
in the introduction. The H1 and the LPS data show no $Q^2$ dependence.
The ZEUS FPC data show a small increase in $A$ at the higher $Q^2$
region.
It should be noted that while for the H1 and LPS data, releasing the
$\xpom$ cut to 0.03 seems to have no effect, the deviation of the ZEUS
FPC data from a flat dependence increases from a 2.4 standard deviation
(s.d.) to a 4.2 s.d. effect (not shown).

\begin{figure}[h]
\includegraphics*[height=.15\textheight]{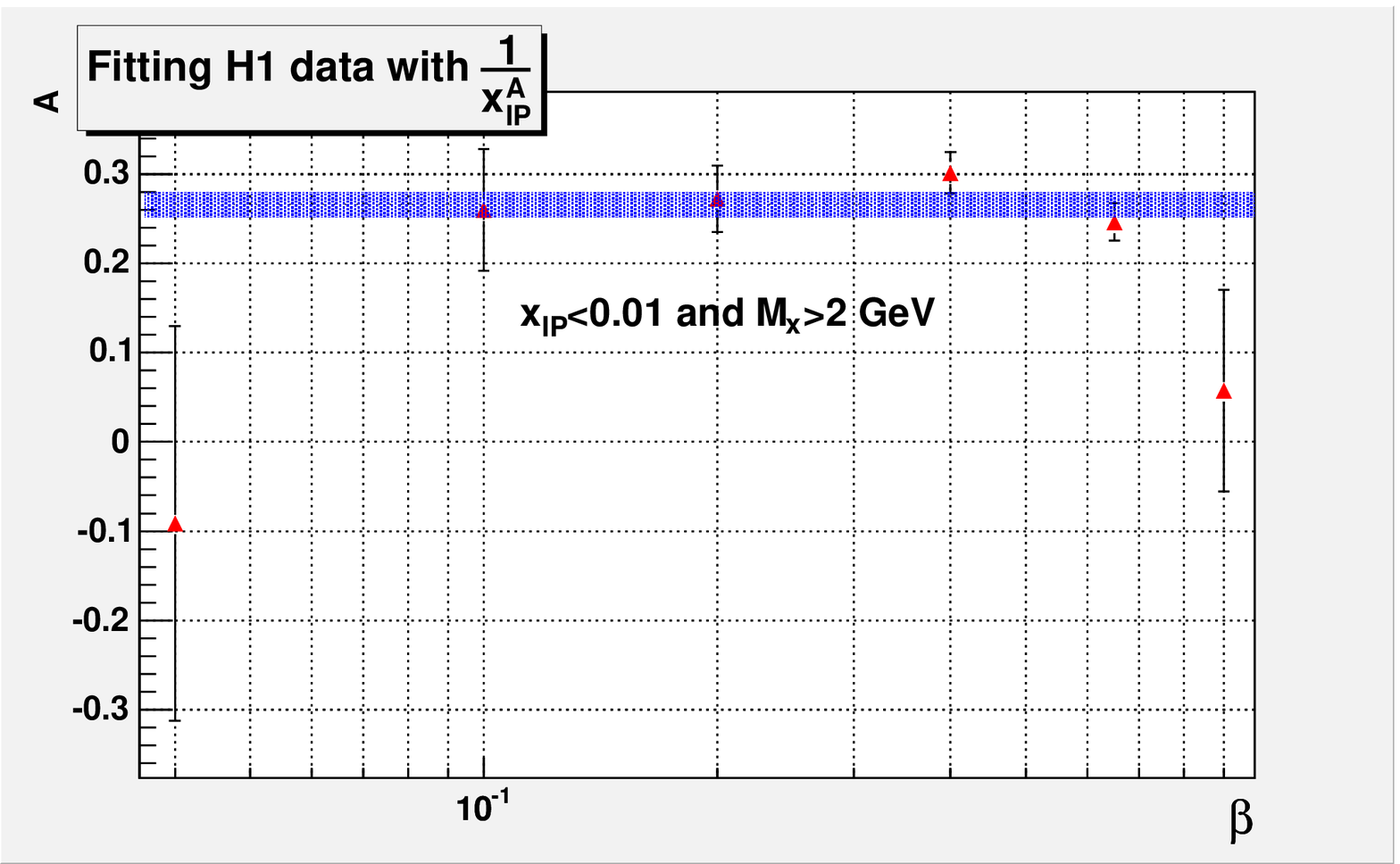}
\includegraphics*[height=.15\textheight]{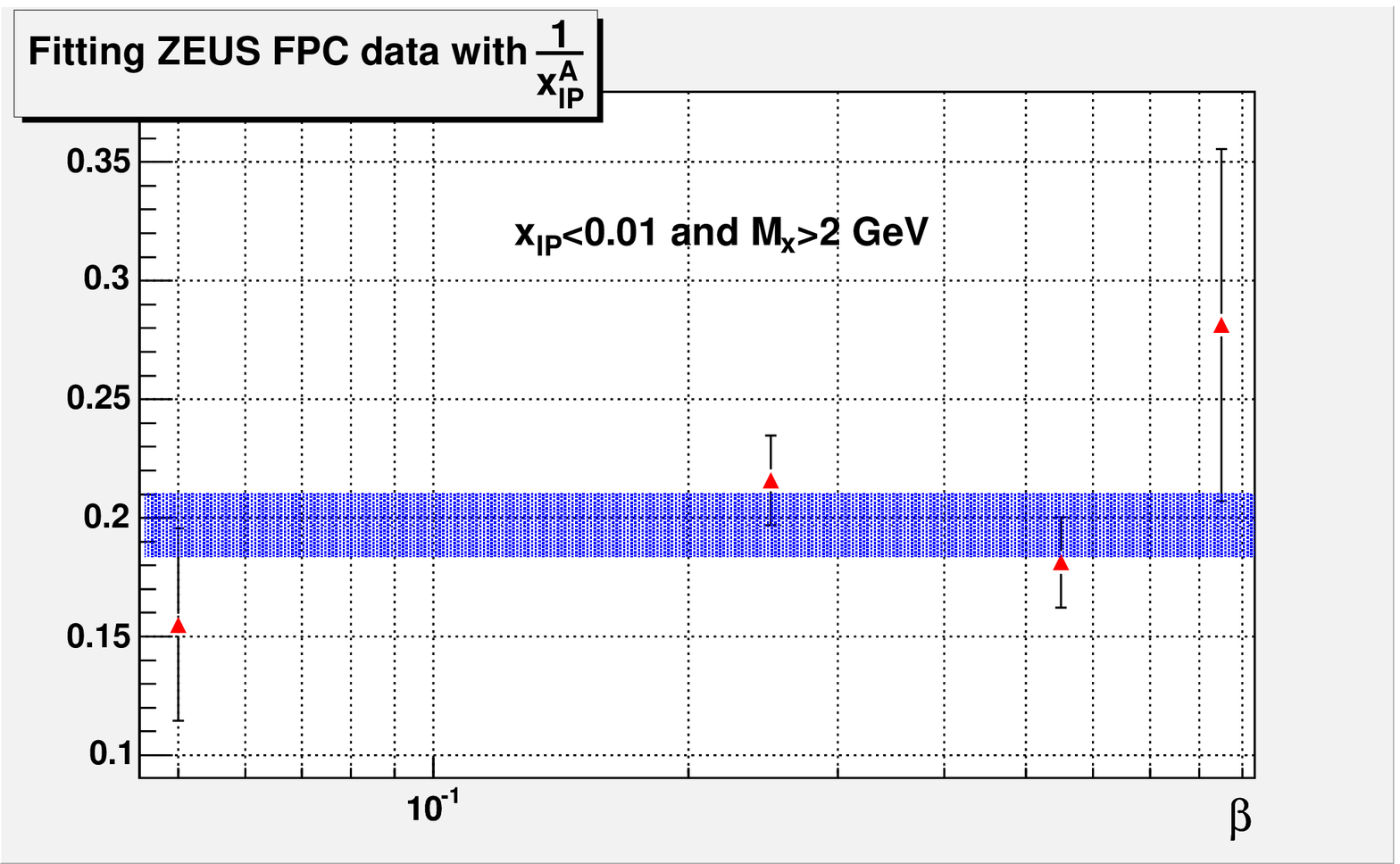}
\includegraphics*[height=.15\textheight]{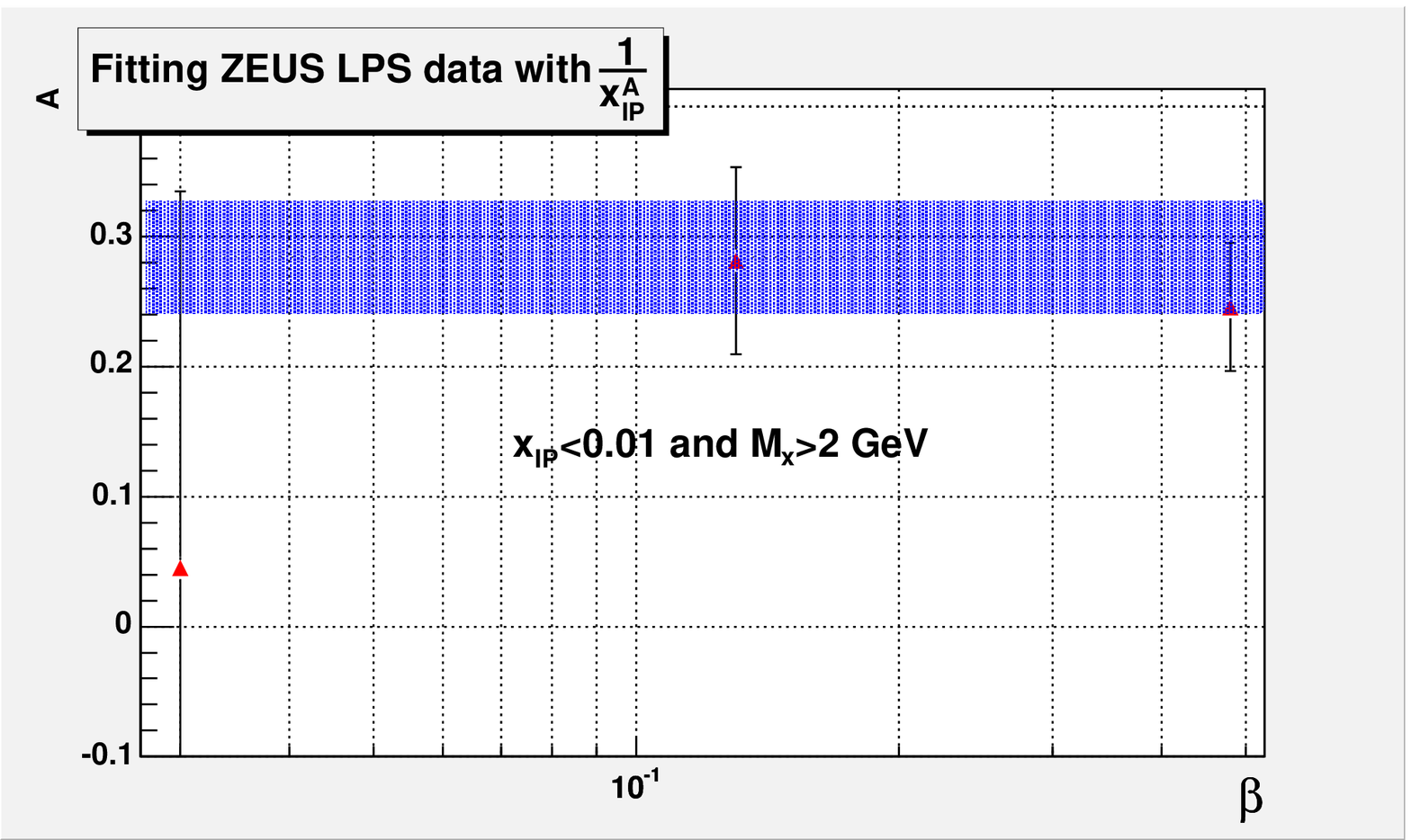}
\caption{$A$ as a function of $\beta$ for $\xpom <$ 0.001 and $M_X >$ 2 GeV, 
for the three data sets, as indicated in the figure. The line corresponds to 
a fit over the whole $\beta$ region}
\label{fig:rf-beta}
\end{figure}

The $\beta$ dependence of $A$ is shown in figure~\ref{fig:rf-beta}.
All three data sets seem to show no $\beta$ dependence, within the errors
of the data. Note however, that by releasing the $\xpom$ cut to higher
values, a strong dependence of the flux on $\beta$ is observed (not
shown).

We thus conclude that for $\xpom <$ 0.01, the Pomeron flux seems to be
independent of $Q^2$ and of $\beta$ and thus the Regge factorization
hypothesis holds.

\section{NLO QCD fits}

\begin{figure}[h]
\includegraphics*[height=.22\textheight]{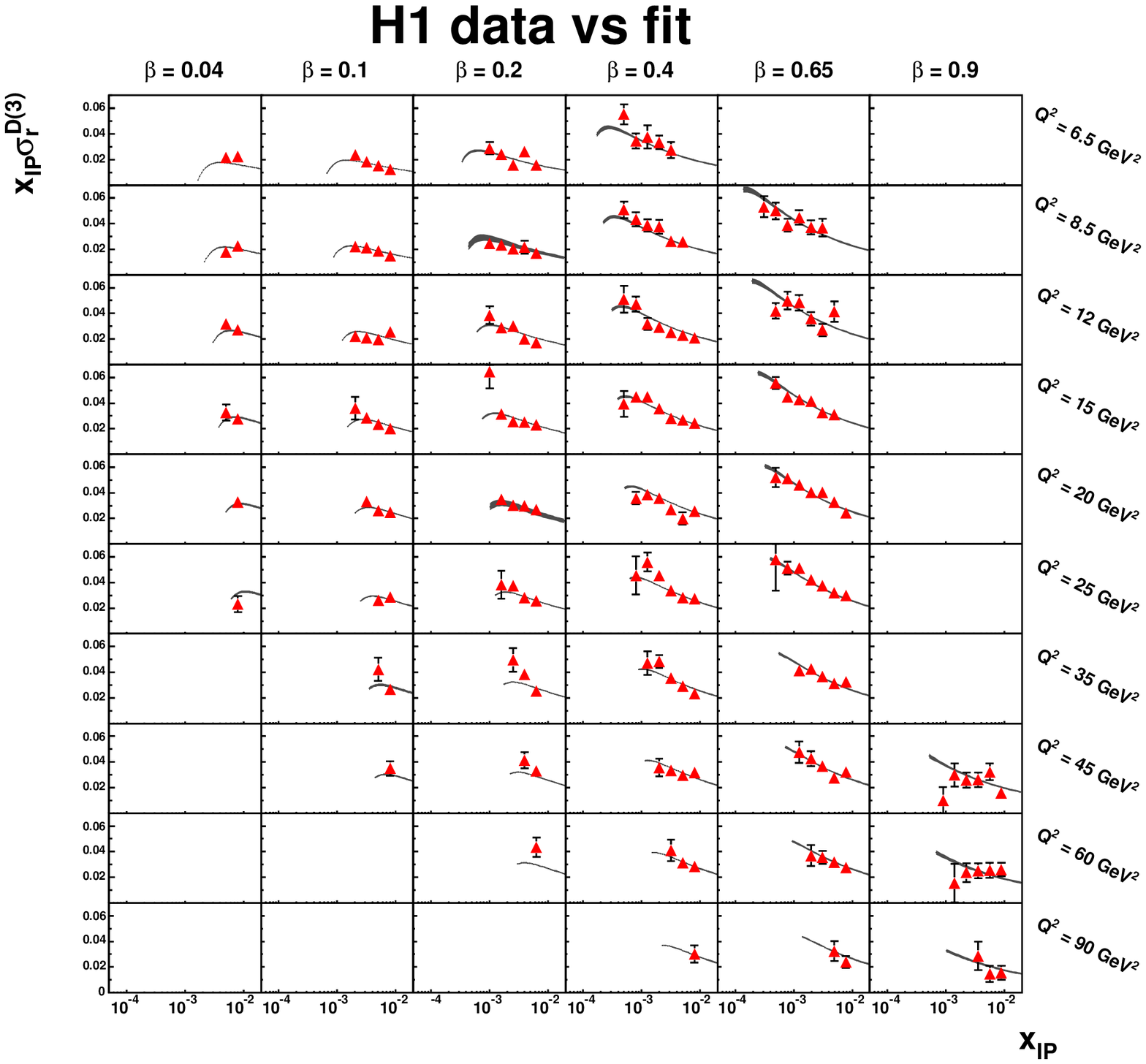}
\includegraphics*[height=.22\textheight]{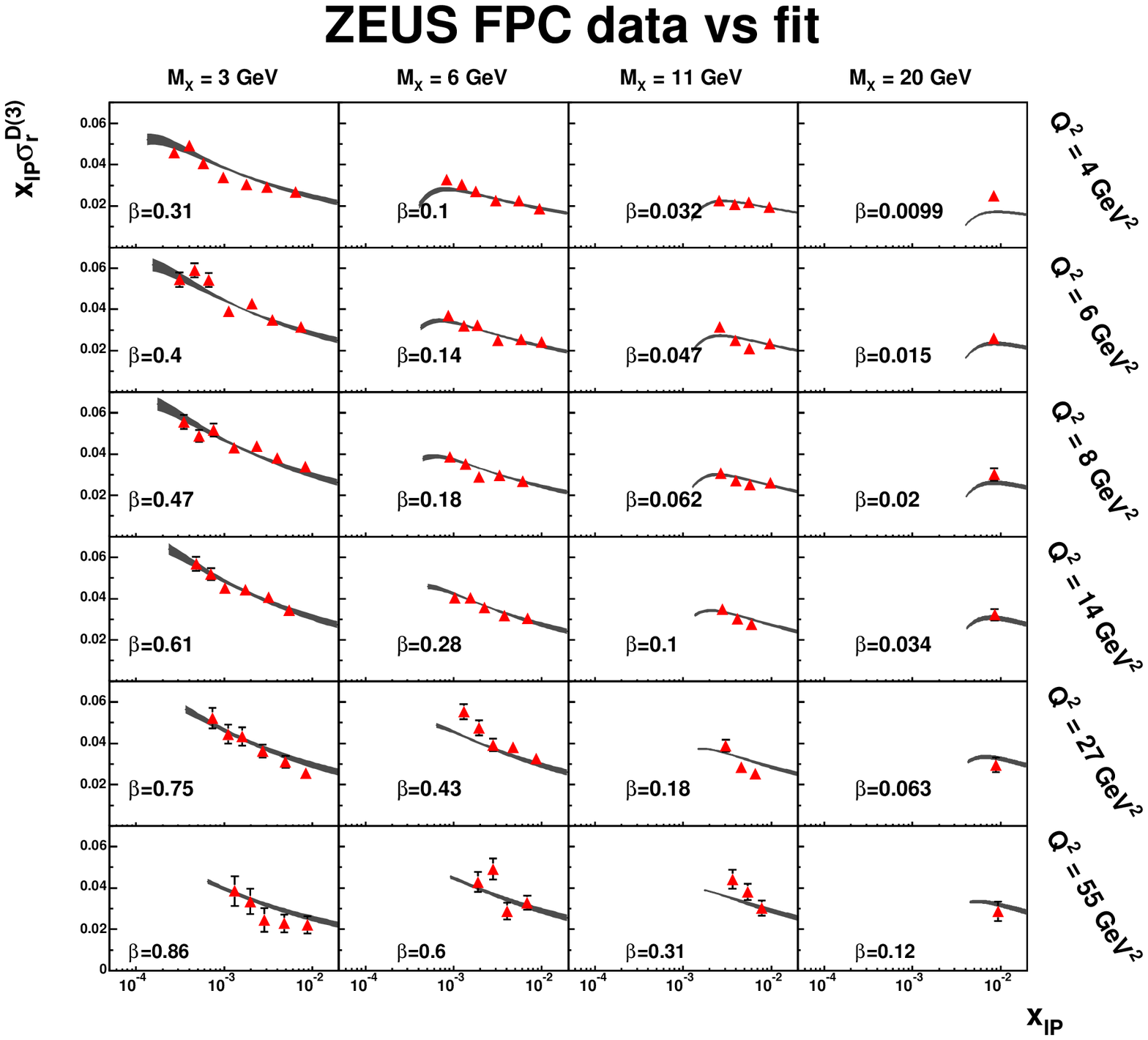}
\includegraphics*[height=.22\textheight]{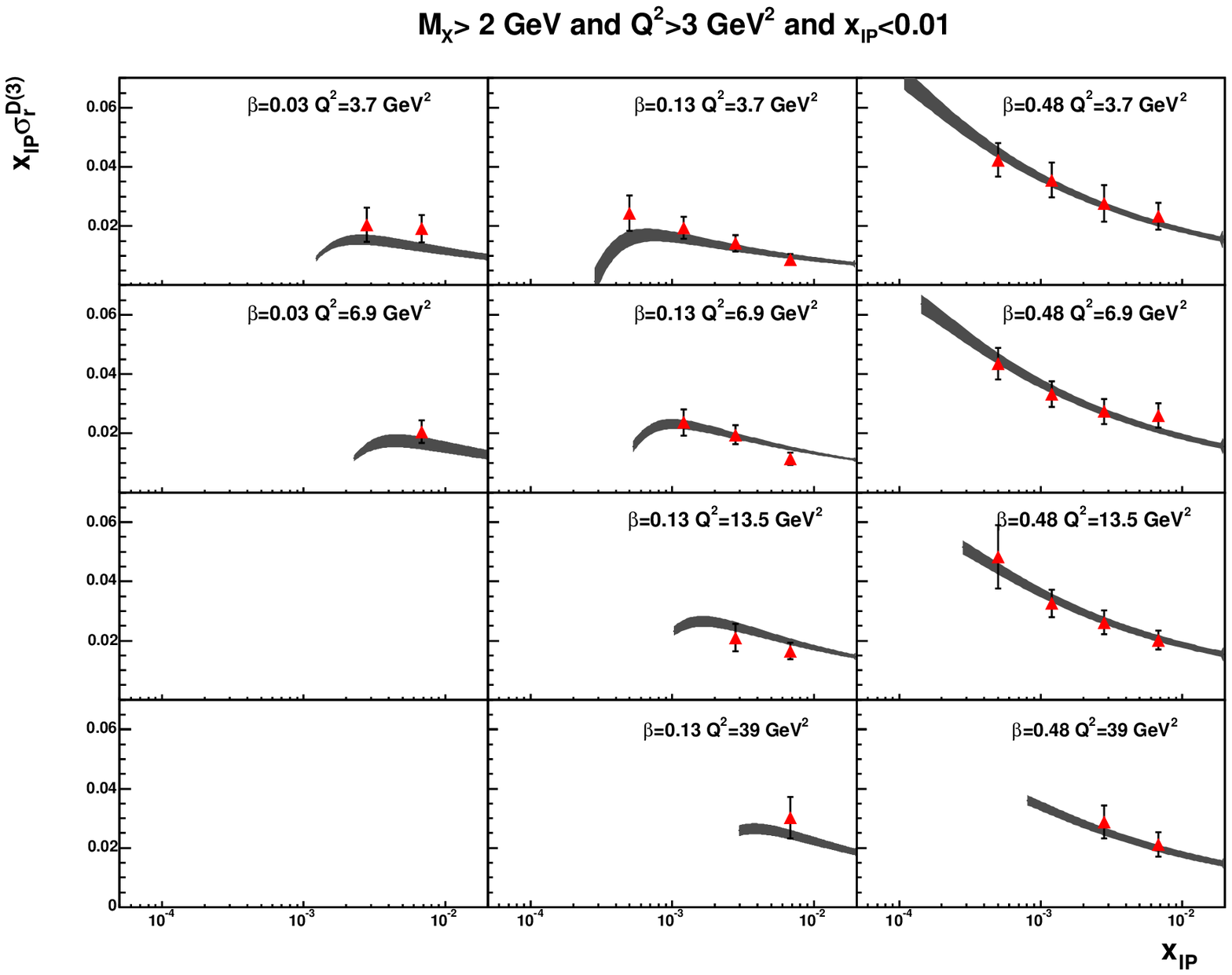}
  \caption{The diffractive reduced cross section of the proton multiplies by 
$\xpom$, as a function of $\xpom$ for the different data sets (the most right 
plot is for the LPS data) in different bins of $Q^2$ and $\beta$, as indicated
 in the figure. The bands are the results of the fits including uncertainties.
}
\label{fig:fits}
\end{figure}

We parameterized the parton distribution functions of the Pomeron at
$Q^2_0$ = 3 GeV$^2$ in a simple form of $Ax^b(1-x)^c$ for $u$ and $d$
quarks (and anti-quarks) and set all other quarks to zero at the
initial scale. The gluon distribution was also assumed to have the
same mathematical form. We thus had 3 parameters for quarks, 3 for
gluons and an additional parameter for the flux, expressed in terms of
the Pomeron intercept $\apom$(0). Each data set was fitted to 7
parameters and a good fit was achieved for each. The H1 and ZEUS FPC
had $\chi^2$/df $\approx$ 1, while for the LPS data, the obtained
value was 0.5. The data together with the results of the fits are
shown in figure~\ref{fig:fits}.
The following values were obtained for $\apom$(0), for each of the
three data sets:
$\apom(0) = 1.138 \pm 0.011, \mbox{for the ZEUS FPC data},
\apom(0) = 1.189 \pm 0.020, \mbox{for the ZEUS LPS data},
\apom(0) = 1.178 \pm 0.007, \mbox{for the H1 data}.$

\begin{figure}[h]
\includegraphics*[height=.4\textheight]{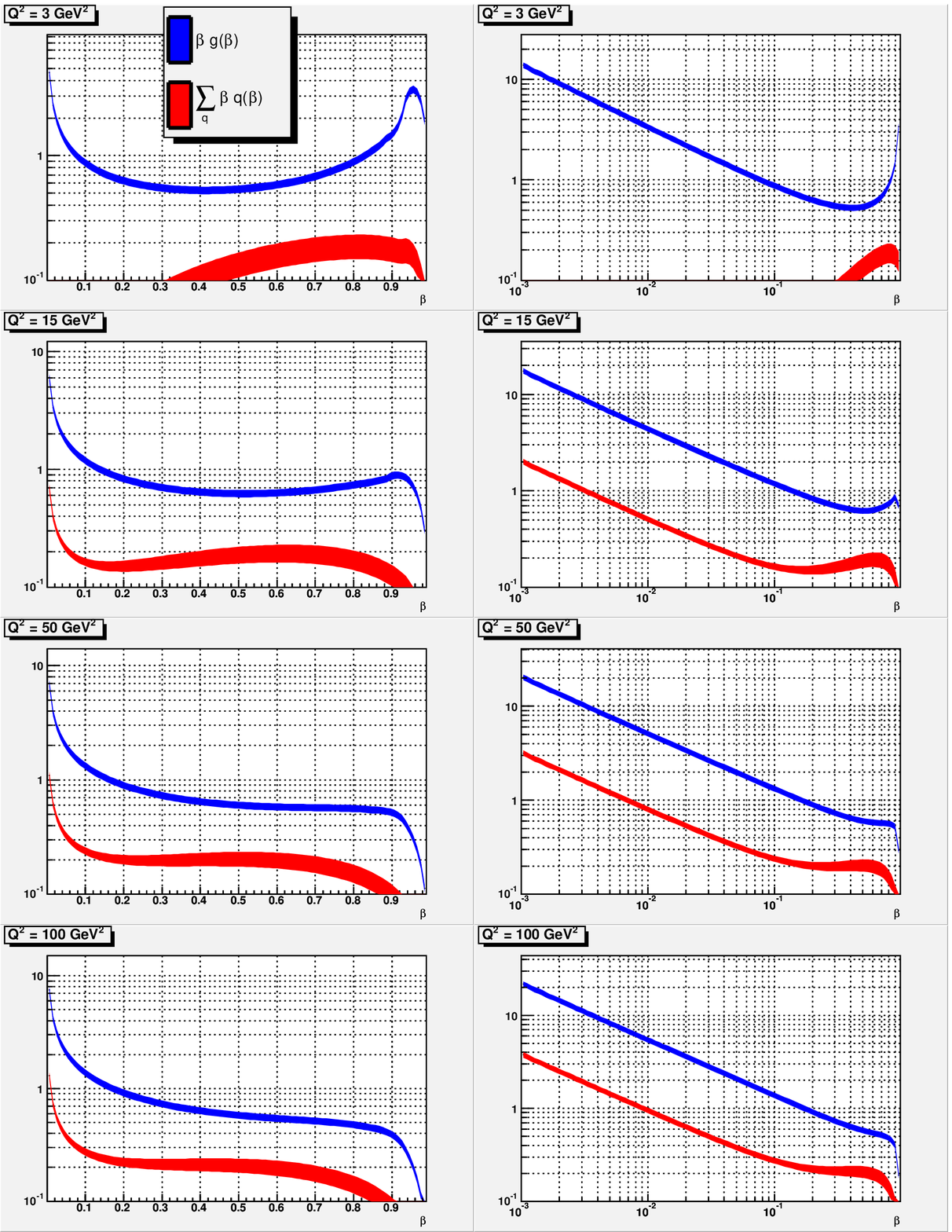}
\includegraphics*[height=.4\textheight]{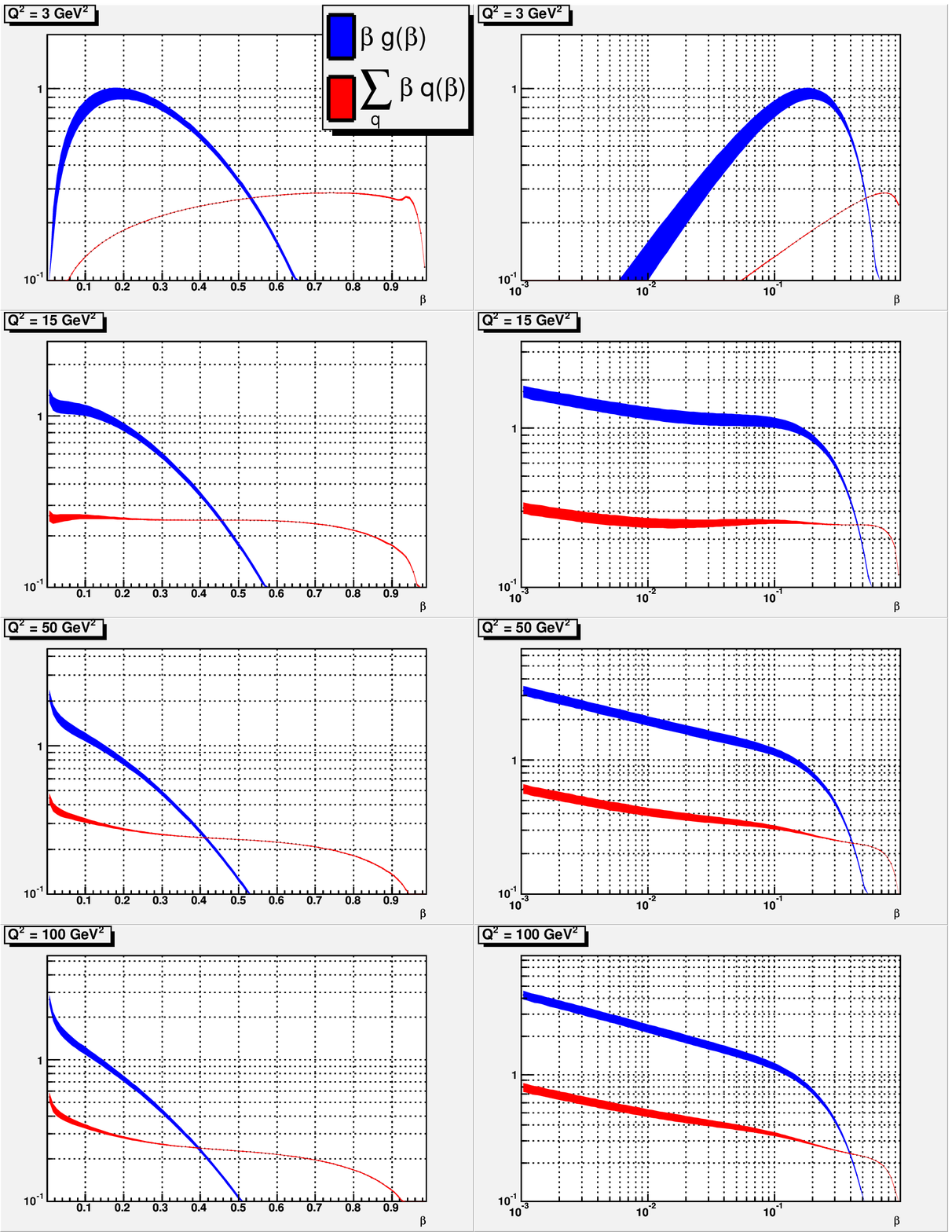}
  \caption{Quark and gluon pdfs of the Pomeron as obtained from the H1 data 
fit (left two figures) and from the ZEUS FPC data fit (two rightmost figures)
 as a function of $\beta$, at different values of $Q^2$.}
\label{fig:pdfs}
\end{figure}

The parton distribution functions are shown in figure~\ref{fig:pdfs}
for the H1 and the ZEUS FPC data points. Because of the limited
$\beta$ range covered by the LPS data, the resulting pdfs
uncertainties are large and are not shown here. For the H1 fit one
sees the dominance of the gluons in all the $\beta$ range. For the
ZEUS FPC data, the quark constituent of the Pomeron dominates at high
$\beta$ while gluons dominate at low $\beta$.
We can quantify this by calculating the Pomeron momentum carried by
the gluons. Using the fit results one gets for the H1 data 80-90\%.
while for the ZEUS FPC data, 55-65\%.

\section{ Comparison of the data sets}
One way of checking the compatibility of all three data sets is to
make an overall fit for the whole data sample. Since the coverage of
the $\beta$ range in the LPS data is limited, we compare only the H1
and the ZEUS PC data. A fit with a relative overall scaling factor of
the two data sets failed. Using the fit results of one data sets
superimposed on the other shows that the fit can describe some
kinematic regions, while failing in other bins. This leads to the
conclusion that there seems to be some incompatibility between the two
data sets.

\section{Probability of diffraction}

\begin{figure}[h]
\includegraphics*[height=.4\textheight]{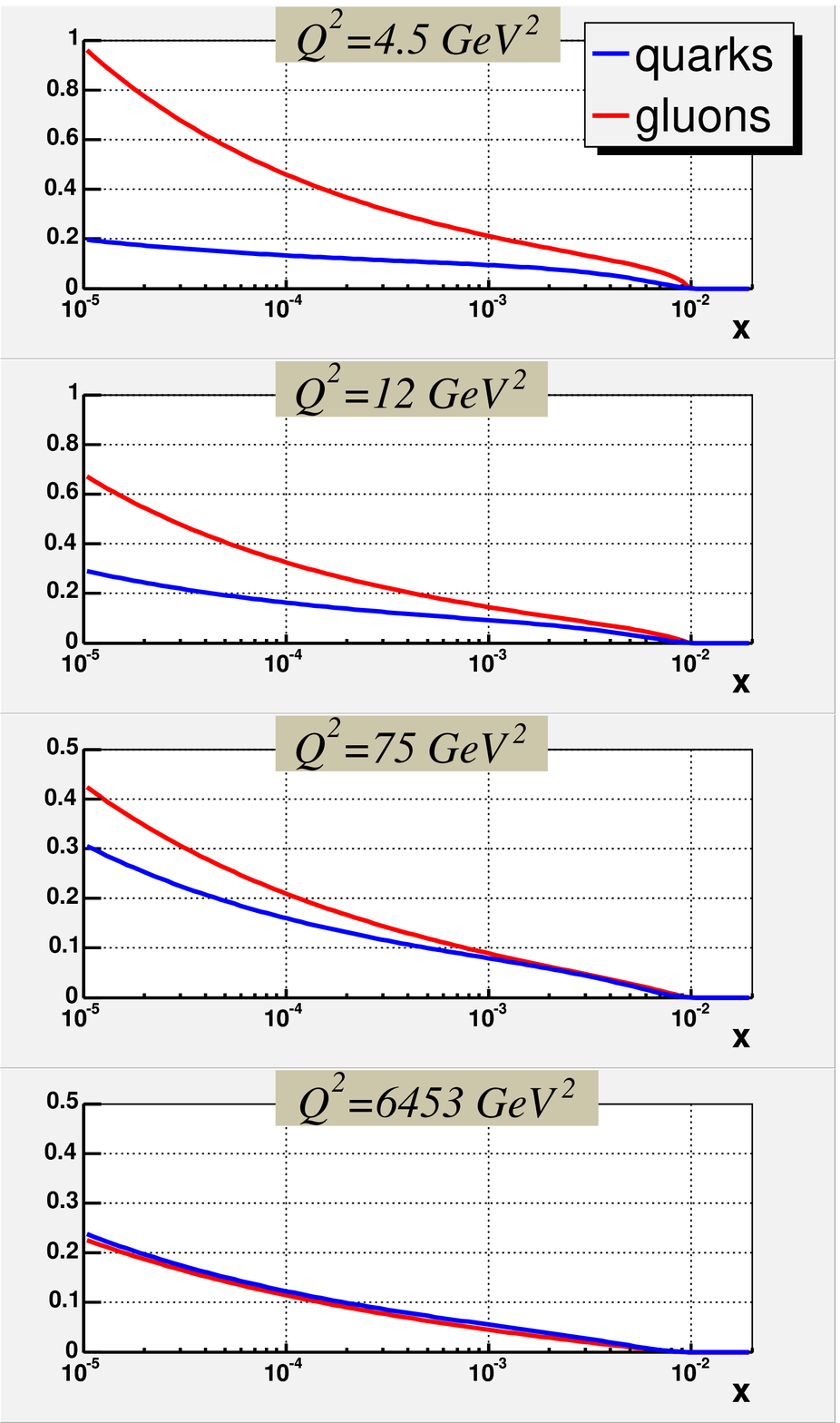}
\includegraphics*[height=.4\textheight]{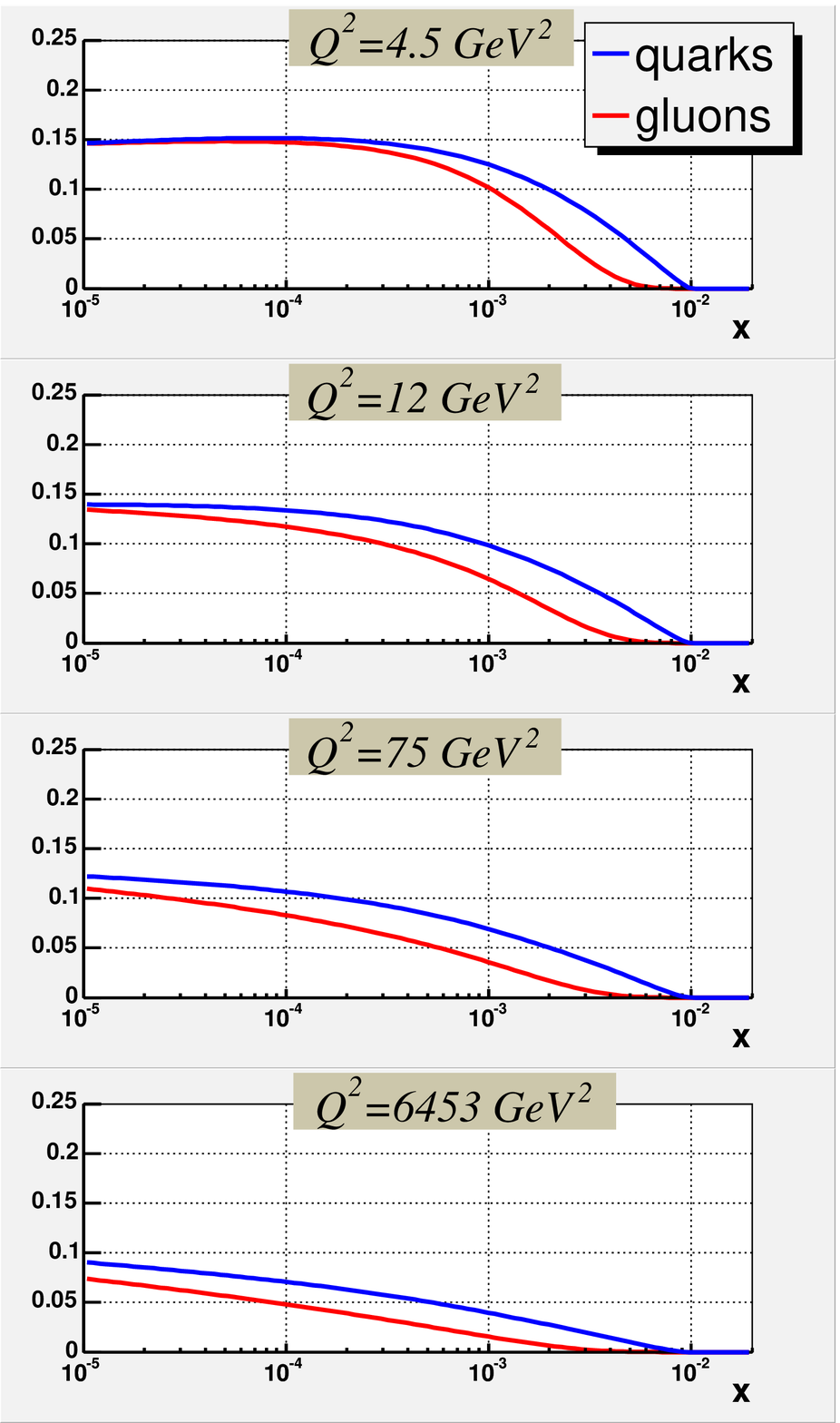}
  \caption{Probability of diffraction as a function of $x$, at different 
values of $Q^2$, calculated from the results of the H1 data fit (left figure)
 and from those of the ZEUS FPC data fit (right figure).}
\label{fig:prob}
\end{figure}

It is of interest to calculate the probability that a certain parton
is produced in a diffractive process~\cite{FS}. The probability of
diffraction on quarks and gluons, as a function of Bjorken $x$ at
different values of $Q^2$ are shown in figure~\ref{fig:prob}, using
the results of the H1 and the ZEUS FPC data fits. 
The ZEUS FPC data shows that throughout the whole kinematic range
shown in the figures, the probability for diffraction is not bigger
than 0.15, far from the Pumplin~\cite{pumplin} limit of 0.5. This is
not the case for the H1 data for which, at small $x$ and low $Q^2$,
the probability of diffraction induced by gluons becomes greater than 0.5 and
thus unphysical. Note however that the results for $x < 2 \cdot
10^{-4}$ are in a region where H1 has no data and thus the calculated
probability in this region is an extrapolation based on the fit
parameters. In order to get physical results, some process, like
saturation, must lower the expected value.


\begin{theacknowledgments}
  We would like to thank Prof. John Collins for providing the program
  to calculate the NLO QCD equations for the diffractive data. This
  work was supported in part by the Israel Science Foundation (ISF).
\end{theacknowledgments}






\end{document}